\documentclass[amsmath,amssymb,aps,twocolumn,showkeys,prl,floatfix]{revtex4-1}

\usepackage{mathrsfs}
\usepackage{amsmath}
\usepackage{amssymb}
\usepackage{color}
\usepackage{graphicx}	% Include figure files
\usepackage{bm}% bold math
\usepackage{ulem} % texte barré

\newcommand{\be}{\begin{equation}}
\newcommand{\ee}{\end{equation}}
\newcommand{\bef}{\begin{figure}}
\newcommand{\eef}{\end{figure}}
\newcommand{\bea}{\begin{eqnarray}}
\newcommand{\eea}{\end{eqnarray}}

\begin{document}
\title{Nanoparticles actively fragment armored droplets}
\author{Fran\c cois Sicard$^{1}$}
\thanks{Corresponding author: \texttt{francois.sicard@free.fr}.}
\author{Jhoan Toro-Mendoza$^{2}$}
\author{Alberto Striolo$^{3}$}
\affiliation{$^1$ Department of Chemistry, King's College London, SE1 1DB London, United Kingdom}
\affiliation{$^2$ Instituto Venezolano de Investigaciones Cientificas, Centro de Estudios Interdisciplinarios de la Fisica, Caracas, Venezuela}
\affiliation{$^3$ Department of Chemical Engineering, University College London, WC1E 7JE London, United Kingdom}

%\date{\today}
%	     
\begin{abstract}
Understanding the complexity of fragmentation processes is essential for regulating intercellular 
communication in mechanistic biology and developing novel bottom-up approaches in a large range of 
multiphase flow processes. In this context, self-fragmentation proceeds without any external mechanical 
energy input allowing one to create efficiently micro- and nanodroplets. Here we examine self-fragmentation 
in emulsion nanodroplets stabilized by solid particles with different surface features. Mesoscopic modelling 
and accelerated dynamics simulations allow us to overcome the limitations of atomistic simulations 
and offer detailed insight into the interplay between the evolution of the droplet shape and the particle 
finite-sized effects at the interface. We show that finite-sized nanoparticles play an \textit{active} role 
in the necking breakup, behaving like nano-scale razors, and affect strongly the thermodynamic properties 
of the system. The role played by the particles during self-fragmentation might be of relevance to 
multifunctional biomaterial design and tuning of signaling pathways in mechanistic biology.
\end{abstract}

%\keywords{self-fragmentation, emulsion droplet, nanoparticle, free-energy, accelerated dynamics simulation}

\maketitle

%\section{Introduction}
%
Emulsion droplets are metastable dispersions comprised of two immiscible fluids such as 
water and oil~\cite{1999-COCIS-Leal-Poulin,2007-ACIS-Shu-vandenBerg}. The associated surface tension 
forces them into a spherical shape to minimize the free-energy. 
To decrease the latter and stabilize the emulsion droplets a surface-active agent can be 
added~\cite{2002-COCIS-Binks,2007-Langmuir-Binks-Desforges}. Pickering emulsions are stabilized by 
the incorporation of particles~\cite{1907-JCS-Pickering,2002-COCIS-Binks}. 
For instance, emulsion droplets can serve as ideal compartments for reactions catalysed by 
nanoparticles (NPs) attached at the oil-water interfaces~\cite{2017-PPSC-Qu-He}, 
can be used as drug-delivery vehicles~\cite{2009-IJP-Frelichowska-Chevalier}, 
sensors~\cite{2016-RSCAdv-Pan-Tang} and templates for the fabrication of advanced functional 
materials~\cite{2011-JMC-Agrawal-Stamm}. 
The characteristics of Pickering emulsions pose a number of intriguing fundamental physical 
questions including a thorough understanding of the perennial lack of detail about how particles 
arrange at the liquid/liquid interface. Other not completely answered questions include particle 
effects on interfacial tension~\cite{2006-ACIS-Miller-Michel},  
layering~\cite{2015-Langmuir-Razavi-Tu}, buckling~\cite{2017-Nanoscale-Sicard-Striolo} 
and droplet bridging~\cite{2018-SM-Bizmark-Ioannidis}.
Interestingly, emulsion droplets show some relevant characteristics and qualities of living systems 
that could make them proxies for artificial life. They provide an experimental framework for synthetic 
biology that is different from other protocell model systems such as vesicles offering distinct 
advantages~\cite{2011-EPL-Ichii-Yomo,2013-CPC-Caschera-Hanczyc,2017-NaturePhys-Zwicker-Julicher}. 
For instance, the apparent similarity in surface properties between inorganic NPs and globular 
proteins ~\cite{2010-Science-Kotov} and the fluid dynamical properties of droplets could be combined 
with different chemistries to target applications and exploration of biological scenarios not easily 
achievable with other supramolecular platforms~\cite{2014-Life-Hanczyc}, including the mechanical biology 
of enveloped viruses such as  the rapidly spreading Zika virus~\cite{2016-JGID-Sikka-Papadimos,2018-Structure-Sevvana-Rossmann,2018-ScienceNews-Saey}.

In the presence of colloidal particles, the stability of emulsions against fragmentation under 
ultrasonication~\cite{2011-IJDD-Ramisetty-Shyamsunder,2012-IECR-Lad-Murthy,2018-NatureComm-Huerre-Garbin} 
or shear stress~\cite{2003-EL-Mabille-Schmitt,2011-Langmuir-Mulligan-Rothstein} 
has been a subject of strong interest either at the molecular or mesoscopic level. 
However, despite the vast interest in particle-laden interfaces, less is know about the self-fragmentation 
mechanism, \textit{i.e} fragmentation without any external mechanical energy input.  
Tcholakova \textit{et al.}~\cite{2017-NatureComm-Tcholakova-Smoukov} studied self-emulsification 
process via cooling-heating cycles causing repeating breakup of droplets to higher-energy states. 
However, this approach relies on small changes in temperature affecting the 
spontaneous curvature due to surfactant thermally-activated modification. 
We are here interested in self-fragmentation at constant temperature, as it is observed in the proliferation 
of living cells~\cite{2011-EPL-Ichii-Yomo,2013-CPC-Caschera-Hanczyc,2014-Life-Hanczyc,2017-NaturePhys-Zwicker-Julicher}.\\

Due to their inherent limited resolution, direct access to local observables, such as the particles' 
three-phase contact angle distribution, the measure of decrease in interfacial tension with the 
decrease in droplet size, as characterized by the Tolman length~\cite{1949-JCP-Tolman}, or the presence 
of particles remain out of reach. 
These pieces of information can be accessed by numerical simulations. 
All-atom molecular dynamics (MD) simulations have become a widely employed computational technique. 
However, all-atom MD simulations are computationally expensive. Moreover, most phenomena of interest 
can take place on time scales that are orders of magnitude longer than those accessible via all-atom MD. 
Mesoscopic simulations, in which the structural unit is a coarse-grained representation of a large number 
of molecules, and enhanced sampling techniques allow us to overcome these limitations. 
Coarse-grained approaches offer the possibility of answering fundamental questions responsible for 
the collective behaviour of particles anchored at an interface~\cite{2018-Elsevier-Wu-Striolo}.
\begin{figure*}[t]
\includegraphics[width=0.95 \textwidth, angle=-0]{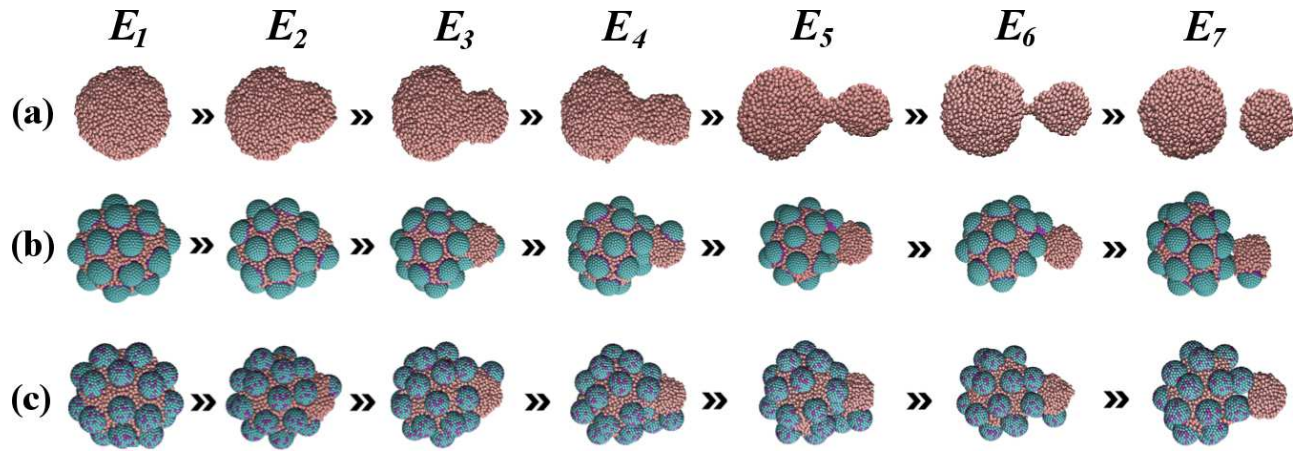}
 \caption{Sequence of simulation snapshots representing the \textit{thermally-assisted} self-fragmentation 
 processes of water in oil nanodroplets in the bare configuration (a) or armored with 24 spherical Janus (b) 
 and homogeneous (c) NPs. The bare and armored nanodroplets are constituted of the same number 
 of water beads, $N_W \approx 3500$.  The same NP surface coverage $\phi \approx 0.9$, as defined 
 in Ref.~\cite{LuuLangmuir2013}, is considered on the armored nanodroplets. The NPs simulated have 
 the same volume, $4/3 \pi a_0^3$, where $a_0 \approx 1.5$ nm. This yields the radius of gyration, 
 $R_{GYR} = (3.764 \pm 0.004)$ nm, and $R_{GYR} = (3.953 \pm 0.008)$ nm for bare and armored 
 nanodroplets, respectively. Cyan and purple spheres represent polar and apolar beads. Pink spheres represent 
 water beads. The oil beads surrounding the droplets are not shown for clarity. The number of  contacts, 
 $N_C$ (see Methods), decreases from left to right, with $E_1$ corresponding to the \textit{original} 
 configuration and $E_7$ to the final configuration obtained within the accelerated dynamics framework.}
\label{fig1}
\end{figure*}

We employ here Dissipative Particle Dynamics (DPD) as a mesoscopic simulation 
method~\cite{1997-JCP-Groot-Warren} along with accelerated dynamics simulations~\cite{2004-JCP-Hamelberg-McCammon} 
to study accurately the free-energy landscapes and the necking breakup mechanisms associated 
with \textit{thermally-assisted} self-fragmentation of model water nanodroplets coated with spherical 
NPs and immersed in an organic solvent. The scale temperature in the DPD framework is equivalent 
to $298.73$ K.
The procedure and parametrisation details are described in the Supplementary Information (SI). 
The particles are of two types: Janus (particles whose surface shows two distinct wetting properties) 
and homogeneous (particles with one surface characteristic). They 
are chosen so that the initial three-phase contact angles ($\approx 90^\circ$) result in maximum 
adsorption energy~\cite{2000-Langmuir-Binks-Lumsdon}.
To study the thermodynamic properties of the self-fragmentation mechanism and the role played 
by the finite-sized particles, we combined several enhanced sampling frameworks, metadynamics 
(metaD)~\cite{2002-PNAS-Laio-Parrinello}, umbrella sampling (US)~\cite{2011-WileyIRCMS-Kastner}, 
and adiabatic biased molecular dynamics (ABMD)~\cite{1999-JCP-Marchi-Ballone,2016-FD-Sicard-Striolo}. 
To take into account the inherent fluid nature of the system, we designed a \textit{dynamic} collective 
variable (CV) allowing to bias unequivocally the dynamics of the system based on dynamical labelling of 
the fluid molecules composing the emulsion droplet as well as the \textit{mother} and the \textit{daughter} 
emulsion droplets resulting from self-fragmentation (see Methods). We consider throughout this study 
the same number of fluid molecules constituting the bare or armored emulsion nanodroplets, and the same 
NP density on the armored nanodroplets.\\

In Fig.~\ref{fig1} we show representative snapshots obtained in the accelerated dynamics simulations 
for the bare emulsion nanodroplet (panel a) and a system stabilized with either Janus (panel b) 
or homogenous (panel c) NPs.
Starting with the spherical bare droplet (configuration $E_1$ in Fig.~\ref{fig1}a), 
the system fragments with the formation of a liquid bridge whose shape 
approximately describes an evolving parabola ($E_2$ to $E_6$). The curvature of the liquid bridge increases continuously 
along the fragmentation process until \textit{mother} and \textit{daughter} nanodroplets separate, 
resulting in two distinct emulsion nanodroplets with radius of gyration 
$R^{(m)}_{GYR} = (3.521 \pm 0.004)$ nm and $R^{(d)}_{GYR} = (2.188 \pm 0.007)$ nm, respectively ($E_7$).
\begin{figure}[b]
\includegraphics[width=0.8 \columnwidth, angle=-0]{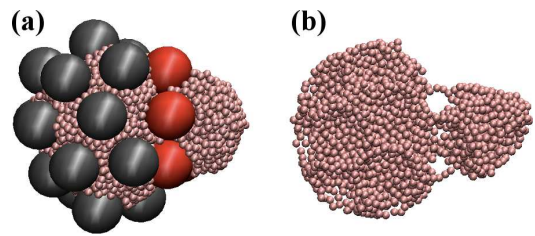}
 \caption{Simulation snapshot obtained during the \textit{thermally-assisted} self-fragmentation process 
 of the emulsion droplet armored with Janus NPs. (a) The edging NPs separating \textit{mother} and \textit{daughter} droplets are coloured in red. The particles which are not involved in 
 the fragmentation process are coloured in grey. (b) Illustration of the filament bridge formed in 
 the necking breakup mechanism when the particles in panel (a)  are not shown. Pink spheres 
 represent water beads. The oil beads are not shown for clarity.}
\label{fig2}
\end{figure}
\begin{figure*}[t]
\includegraphics[width=0.99 \textwidth, angle=-0]{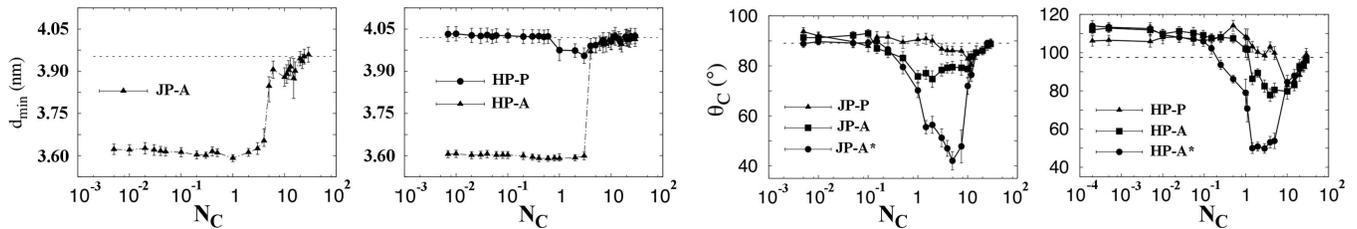}
 \caption{Evolution of the representative minimal distances $d_{min}$ (left panels) measured between 
 the edging NPs and the representative three-phase contact angles $\theta_{C}$ (right panels) of Janus (JP) 
 and homogeneous (HP) edging NPs as a function of the number of contacts $N_C$ between 
 \textit{mother} and \textit{daughter}  droplets (see Methods). The letters $P$ and $A$ are used to differentiate 
 the NPs which play a \textit{passive} and an \textit{active} role in the breakup mechanism, 
 respectively. The statistical errors are estimated as one standard deviation from the average as obtained 
 for equilibrated trajectories. The values  $d_{min}$ and $\theta_{C}$  of a representative NP far from 
 the fragmentation area  are plotted in dashed line for reference.}
\label{fig3}
\end{figure*}
The fragmentation process is fundamentally different when the emulsion droplet is stabilized with 
either Janus or homogeneous NPs (cf. Figs.~\ref{fig1}b and \ref{fig1}c). As the particles remain strongly 
adsorbed at the interface during fragmentation, forming a close-packed monolayer, the emulsion nanodroplet 
fragments forming first a liquid bridge similar to the one observed in the bare droplet (configuration $E_2$ 
in Fig.~\ref{fig1}).
Progressively, the particles located at the fragmentation edge \textit{actively} control the process. 
They decrease the number of contacts between \textit{mother} and \textit{daughter} droplets. 
As shown in Fig.~\ref{fig2}, these edging NPs control the shape of the bridge between the 2 droplets, 
behaving like nano-scale razors, and transform the liquid bridge in bridging filaments. Finally, the 
bridging filaments vanish and \textit{mother} and \textit{daughter} emulsion droplets separate.
 
When the armored nanodroplet is coated with Janus NPs (panel b), their hydrophobic regions interact with the  \textit{daughter} droplet, pushing it further away 
from the \textit{mother} droplet. The final configuration is fundamentally different when the water nanodroplet 
is stabilized with homogeneous NPs (panel c). Their specific feature allows the edging NPs to adsorb both at 
the \textit{mother} and \textit{daughter} droplet interfaces once they are separated, resulting in 
bridging~\cite{2016-FD-Sicard-Striolo}. Remarkably, as the distance between  \textit{mother} 
and \textit{daughter} droplets is not sufficiently large, the fragmented emulsion droplets can coalesce.\\

The  breakup mechanism just discussed is quantitatively investigated in Fig.~\ref{fig3} (left panels), 
where we show the evolution of a representative set of minimal distances, $d_{min}$, measured between 
the edging NPs separating \textit{mother} and \textit{daughter} nanodroplets
as a function of the number of contacts, $N_C$, between the water molecules in each droplets. 
The initial distributions of the minimal distances can be described with Gaussian distributions 
for both Janus (J) and homogeneous (H) NPs. The values of the respective means, $\mu^J$ and $\mu^H$, 
and standard deviations, $\sigma^J$ and $\sigma^H$, differ due to the NP features. We obtain 
$\mu^J = 3.96$ nm and $\mu^H = 4.02$ nm, and $\sigma^J = 0.03$ nm and $\sigma^H = 0.02$ nm.
As the number of contacts $N_C$ between \textit{mother} and \textit{daughter} droplets coated with Janus NPs
decreases, the minimal distances between the edging NPs show two distinct regimes separated with a transition 
at $N_C^* \approx 5$. When $N_C > N_C^*$, $d_{min}$ decreases slowly 
as the liquid bridge forms, similar to the one observed in the bare droplet. When $N_C \approx N_C^*$, $d_{min}$ 
shows a significant jump which is characteristic of the transition from the liquid bridge to the bridging filaments shown 
in Fig.~\ref{fig2}. When $N_C < N_C^*$, $d_{min}$ decreases continuously until the bridging filaments vanish 
at $N_C^\dag \approx 1$ and \textit{mother} and \textit{daughter} droplets 
separate. For $N_C < N_C^\dag$, $d_{min}$ increases slightly until a plateau is reached, which is characteristic 
of the local rearrangement of the edging NPs near the fragmentation area. 
The evolution of the system is qualitatively similar when homogeneous NPs are present, albeit some subtle differences are present due to the NP features. As shown in Fig.~\ref{fig3}, two distinct behaviours are 
observed. Unlike Janus edging NPs, certain homogeneous NPs do not present a distance $d_{min}$ 
showing a significant jump when $N_C = N_C^*$ but follow a relatively small and continuous decrease 
for $N_C > N_C^\dag$. The rest of the evolution is similar to the one observed for Janus NPs, with 
the continuous increase of $d_{min}$ to the plateau characteristic of the local equilibrium rearrangement 
of the edging NPs.\\

This analysis can be completed by quantifying the evolution of the three-phase contact angles, $\theta_C$, 
of the edging NPs, as shown in Fig.~\ref{fig3} (right panels). The initial distribution can be described 
with Gaussian distributions for both Janus (J) and homogeneous (H) NPs, with respective means and 
standard deviations $\mu^J = 89.1^\circ$ and $\mu^H = 97.5^\circ$, and $\sigma^J = 1.6^\circ$ 
and $\sigma^H = 3.5^\circ$.
As the number of contacts $N_C$ between  \textit{mother} and \textit{daughter} droplets 
coated either with Janus or homogeneous NPs decreases, two distinct behaviours emerged, 
which could discriminate the \textit{active} or \textit{passive} role played by the edging NPs 
in the self-fragmentation process. When $N_C > N_C^*$, the three-phase contact angle decreases slowly 
as the liquid bridge forms. 
When $N_C^\dag < N_C < N_C^*$, we observe two different evolutions of the contact angle.
Some NPs  show a continuous and significant decrease of $\theta_C$, with a decrease of 
$\approx 20\%$ (index $A$ in Fig.~\ref{fig3}) and $\approx 50\%$ (index $A^*$ in Fig.~\ref{fig3}). 
This behaviour is representative of the \textit{active} role played by some edging NPs 
which behave like nano-scale razors. Eventually, the bridging filaments vanish 
when $N_C = N_C^\dag$ and  \textit{mother} and \textit{daughter} droplets separate for $N_C > N_C^\dag$. 
This step is characterized by a significant increase of the contact angle until it reaches a plateau 
associated with the local rearrangement of the edging NPs resulting in higher value of the contact angles 
near the fragmentation area. 
In contrast, some edging NPs show an increase of their contact angle when $N_C < N_C^*$, until it reaches 
the plateau, characteristic of their \textit{passive} role during the breakup mechanism (index $P$ in Fig.~\ref{fig3}).\\

\begin{figure}[t]
\includegraphics[width=0.8 \columnwidth, angle=-0]{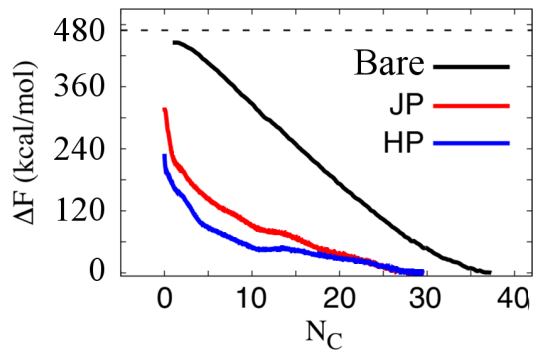}
 \caption{Free-energy profiles associated with \textit{thermally-assisted} self-fragmentation 
 of the bare and armored droplets as a function of the number of contacts $N_C$ between 
 \textit{mother} and \textit{daughter} droplets (see Methods). The Gibbs free-energy calculated 
 without the Tolman correction is shown as a dashed horizontal line. The positions of the global minima 
 of the bare and armored droplets along $N_C$ differ due to the presence of the particles.}
\label{fig4}
\end{figure} 
Finally, building on accelerated dynamics frameworks~\cite{2004-JCP-Hamelberg-McCammon}, 
we assessed the free-energy of self-fragmentation for bare and armored nanodroplets. 
Considering first the bare droplet, we obtained  $\Delta F_B = (445 \pm 3)$ kcal/mol. 
This value can be compared with the expression of the Gibbs free-energy~\cite{2004-Springer-deGenes-Quere}, 
$\Delta F = \gamma~\Delta A$, with $\gamma$ and $\Delta A$ the liquid-liquid interfacial tension 
and the change in interfacial area, respectively. Taking into account the effect of the curvature 
of the emulsion nanodroplet~\cite{1949-JCP-Tolman,2005-JACS-Lei-Zeng}, 
one must consider $\gamma = \gamma_0 / (1+2 \delta / R_S)$ with $\gamma_0$, $R_S$, and $\delta$ denoting 
the planar interfacial tension, the radius of the surface under tension, and the Tolman length, respectively. 
Given the interfacial tension for a planar decane/water interface~\cite{2012-PRE-Fan-Striolo}, 
$\gamma_0 = 51.7 mN.m^{-1}$, we obtained $\Delta F_{0} = (467 \pm 5)$ kcal/mol, which yields the value 
of the Tolman length $\delta \approx 10^{-8}$ cm, in agreement with the original paper of 
Tolman~\cite{1949-JCP-Tolman}. Noticeably, this latter result was directly obtained from our 
simulations and not as a condition imposed where a definition of surface density profiles are needed 
in atomistic descriptions.

\begin{figure}[b]
\includegraphics[width=0.7 \columnwidth, angle=-0]{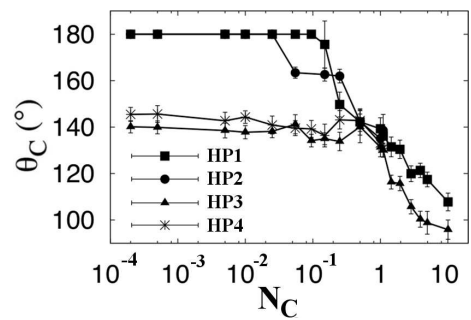}
 \caption{Representative three-phase contact angles $\theta_{C}$ of the homogeneous (HP) edging NPs 
 with respect to the \textit{daughter} droplet as the fragmentation process progresses from
stage E4 to E7 as a function of the continuous number of contacts $N_C$ between \textit{mother} and \textit{daughter} droplets (see Methods). The statistical errors are estimated as one standard deviation 
from the average obtained for equilibrated trajectories.}
\label{fig5}
\end{figure} 
In Fig.~\ref{fig4}, we compared this result with the free-energy of self-fragmentation for the armored 
nanodroplets, $\Delta F_J$ and $\Delta F_H$, respectively, 
also obtained within the accelerated dynamics framework.
As we could expect from the quantitative analysis above, which highlighted the active role played by 
the finite-sized NPs in the  breakup mechanism, we measured 
$\Delta F_J = (316 \pm 7)$ kcal/mol and $\Delta F_H = (227 \pm 3)$ kcal/mol, significantly lower than 
the free-energy of self-fragmentation measured for the bare nanodroplet. 
Furthermore, particle adsorption at the two interfaces when homogeneous NPs are present must be taken 
into account to estimate the fragmentation free-energy. To do so, we measured the three-phase contact angle 
of the edging NPs  and defined the final fragmentation stage as the one where the edging NPs remain adsorbed at a single interface.
In Fig.~\ref{fig5}, we show the evolution of the representative three-phase contact angles of the edging NPs 
with respect to the \textit{daughter} droplet as the fragmentation process progresses from stage $E_4$ to $E_7$,
as shown in Fig.~\ref{fig1}. As the edging particles $HP_1$ and $HP_2$ are fully desorbed when the number of contacts 
$N_C \approx 10^{-2}$, with a three-phase contact angle $\theta_C = 180^\circ$, two particles, $HP_3$ and $HP_4$, 
have a contact angle $\theta_C < 180^\circ$ characteristic of particles partially adsorbed at the \textit{daughter} interface.
In the Pieranski-Binks approach~\cite{2000-Langmuir-Binks-Lumsdon,2011-Langmuir-Wi-Law}, the change 
in energy accompanying the desorption of a spherical particle from the interface to either bulk phase 
can be approximated by $\Delta E = \pi r^2 \gamma_0 (1 \pm \cos \theta)^2$, in which $r$ is the particle 
radius and the plus (minus) sign refers to desorption into oil (water). Even if this expression assumes 
that the oil-water interface remains planar, it yields a rough approximation of the energy at play. 
Considering the contact angles given in Fig.~\ref{fig5}, 
we obtain a correction factor $\Delta \Delta F_H \approx 5$ kcal/mol. 

Interestingly, the difference in chemistry of the spherical NPs had some important impact in the reduction 
of the free-energy barrier. Unlike Janus NPs, which present a preferred orientation at the liquid-liquid 
interface resulting in restricted rotational mobility, homogeneous NPs are characterized by larger 
rotational freedom. This provides the homogeneous NPs with the capability to share easily the interfacial 
area delimiting \textit{mother} and \textit{daughter} nanodroplets during the fragmentation process.
In contrast, when Janus NPs are present, the main energy consumption comes from the \textit{daughter} 
droplet trying to maintain the orientation of the edging NPs, thus increasing the free-energy of 
self-fragmentation.\\

%\section{Conclusion}
%
The physical insights discussed in this letter provide a deeper understanding of the organisation 
of finite-sized NPs at fluid interfaces, and allow us to decipher the \textit{active} or \textit{passive} 
role played by the particles in the self-fragmentation process. 
This information could be useful for a variety of applications including multifunctional biomaterial design 
and tuning of signaling pathways in mechanistic biology, including membrane-bound and membrane-less organelles 
 undergoing apoptosis~\cite{2013-Cells-Bottone-Scovassi,2017-DevBiol-AguileraGomez-Rabouille,2018-Cells-Chen-Stark}.
Apoptosis is a form of programmed cell death that is a highly regulated and controlled biological 
process~\cite{2007-TP-Elmore}, which can alter organelle structure and function~\cite{2013-Cells-Bottone-Scovassi}. 
Its original role is to kill infected, abnormal, or otherwise undesired cells. There is a long list 
of diseases associated with altered cell survival. Increased apoptosis is characteristic of AIDS, 
and neurodegenerative diseases such as Alzheimer's and Parkinson's diseases. Decreased or inhibited 
apoptosis, on the other hand, is a feature of many malignancies, autoimmune disorders, and some viral 
infections~\cite{2001-BMJ-Renehan-Potten}. 
For instance, increasing evidence suggests that cell-derived extracellular vesicles (EV)
produced during apoptosis have important immune regulatory roles relevant across different disease settings~\cite{2018-FI-Caruso-Poon}. The formation of EVs during apoptosis could be a key mechanism of immune 
modulation. 
With most cancer treatments focusing on inducing apoptosis in tumor cells~\cite{2011-JECCR-Wong}, 
it becomes important to understand selective ways to influence cell differentiation and death.
Nanoscale chemistry and topography could act synergistically for better understanding of hidden mechanisms 
of nanomaterial-induced cell behaviors. From this perspective, emulsion droplets can promote more open 
thinking about how non-living matter might self-organize into evolving matter that adapts over time to 
a changing environment.

The extensive simulations discussed above allowed us to decipher  
\textit{thermally-assisted} self-fragmention of armored nanodroplet along with the mechanisms at play 
in the necking breakup process. We showed that finite-sized NPs can play an \textit{active} 
role, behaving like nano-scale razors, during the evolution of the droplet shape and affect strongly the stability 
of the system, resulting in significant reduction of the fragmentation free-energy ranging 
between $150 - 250$ kcal/mol, which is equivalent to the energy released by $20-30$ ATP 
molecules~\cite{1986-JBC-Gajewski-Goldberg}.
%over several hundreds of $k_B T$. 
%
The DPD framework considered in this work would allow extending these results to a range of 
bio-inspired liquid-liquid systems where the adsorption of the particles does not lead to a significant 
deformation of the interface. This is a valid approximation for particles ranging from nanometer to 
micrometer size~\cite{2018-Elsevier-Wu-Striolo}.
These properties might be of relevance for the control and/or tuning of the fragmentation of cell-derived 
extracellular vesicles, which have important immune regulatory roles~\cite{2018-Cells-Chen-Stark}, 
and bacterial membrane vesicles that affect diverse biological processes, including virulence, phage infection, 
and cell-to-cell communication~\cite{2019-NatureRevMicrobiol-Toyofuku-Eberl}.
They might also pave the way for new types of nanoscale platforms in synthetic biology to modify the molecular workings of enveloped viruses such as the rapidly spreading Zika virus whose recent outbreak has been linked 
to microcephaly and Guillain-Barr\'e syndrome~\cite{2016-JGID-Sikka-Papadimos,2018-Structure-Sevvana-Rossmann,2018-ScienceNews-Saey}.

\section*{acknowledgements}

The authors acknowledge V. Garbin, and L. Botto for useful discussions. 
F.S. thanks M. Salvalaglio for fruitful discussion concerning the accelerated dynamics frameworks 
and P. Rousseau, A. Lesne, and M. Barbi for stimulating discussions regarding the biological implications 
of our findings.  Via our membership of the UK’s HEC Materials Chemistry Consortium, which 
is funded by EPSRC (EP/L000202), this work used the ARCHER UK National Supercomputing Service 
(http://www.archer.ac.uk). F.S. acknowledges the support of the UK Engineering and Physical Sciences 
Research Council (EPSRC), under grant number 527889.

\section*{Methods}
\textbf{Dynamical coordination number.} To take into account the inherent fluid nature of the system, 
\textit{i.e.} the absence of covalent interactions between the molecules of the fluids, 
we designed a \textit{dynamic} collective variable (CV) which allowed us to bias unequivocally the dynamics 
of the system based on dynamical labelling of the fluid molecules composing \textit{mother} and 
\textit{daughter} droplets resulting from the self-fragmentation process. 
In this optic, we considered the \texttt{MULTICOLVAR} module of the plugin for free-energy calculation, PLUMED, 
version 2.3~\cite{PLUMED}.
\begin{itemize}
\item We first arbitrarily aligned the principal axis of deformation of the system with the $Z$-Cartesian 
axis of the the DPD simulation box. 
\item We settled the \textit{fragmentation center} as our reference position in the Cartesian space 
using a virtual atom in a fixed position with the \texttt{FIXEDATOM} function available in PLUMED~\cite{PLUMED}. 
\item We used the function \texttt{ZDISTANCES} to calculate the $Z$-components of the vectors 
connecting the fluid molecules constituting the droplet and the \textit{fragmentation center}. 
\item We filtered the distribution of distances obtained with the \texttt{ZDISTANCES} function 
to \textit{dynamically} discriminate the molecules located on the right (left) 
side of the \textit{fragmentation center}. To do so, we used the function \texttt{MFILTER\_MORE} 
(\texttt{MFILTER\_LESS}) implemented in PLUMED~\cite{PLUMED} to create these two dynamic groups. 
\item Thereafter, we fixed the number of molecules constituting \textit{mother} and \textit{daughter} 
droplets. To do so, we used the \texttt{RESTRAINT} function implemented in PLUMED~\cite{PLUMED} 
which can add harmonic and/or linear restaints on specific CVs. We applied the \texttt{RESTRAINT} 
function on two dynamic groups defined with the options \texttt{LESS\_THAN} and \texttt{MORE\_THAN} 
implemented in the \texttt{ZDISTANCES} function.

\item We computed the number of contacts, $N_C$, between the two dynamic groups previously defined with 
\texttt{MFILTER\_MORE} and \texttt{MFILTER\_LESS} using the function \texttt{COORDINATIONNUMBER} 
implemented in PLUMED~\cite{PLUMED}. This variable counts the number of contacts between 
two groups of atoms and is defined as $N_C = \sum_i \sum_j s_{ij}$ with $s_{ij}=1$ if the contact between atoms 
$i$ and $j$ is formed and $s_{ij}=0$ otherwise. To make $N_C$ differentiable $s_{ij}$ 
is replaced with a switching function. We considered the \texttt{RATIONAL} switching function 
$s(r)= \frac{1-(\frac{r-D_0}{R_0})^n}{1-(\frac{r-D_0}{R_0})^m}$, with $n=6$ and $m=12$. 
We choose $D_0=0.6375$ nm, as obtained from the position of the first peak of the 
radial distribution function calculated from the distances between the fluid molecules, 
which correspond to the first coordination shell of the system~\cite{PLUMED}. We then tuned the value 
of the parameter $R_0=0.8625$ nm to allow the \textit{daughter} droplet to separate completely 
from the \textit{mother} droplet during fragmentation.
\item Finally, we defined the continuous number of contacts as the mean value 
of the \texttt{COORDINATIONNUMBER} discussed above. To do so, we used the \texttt{MEAN} option implemented 
in the \texttt{MULTICOLVAR} module available in PLUMED~\cite{PLUMED}.
\end{itemize}

\bibliography{arXiv} %You need to replace "rsc" on this line with the name of your .bib file
\bibliographystyle{apsrev4-1} %the RSC's .bst file

%%%%%%%%%% Merge with supplemental materials %%%%%%%%%%
%\pagebreak
%\widetext
%\begin{center}
%\textbf{\large Nanoparticles actively fragment armored droplets\\
%			   Supplemental Information}
%\end{center}
%%%%%%%%%% Merge with supplemental materials %%%%%%%%%%
%%%%%%%%%% Prefix a "S" to all equations, figures, tables and reset the counter %%%%%%%%%%
%\setcounter{equation}{0}
%\setcounter{figure}{0}
%\setcounter{table}{0}
%\setcounter{page}{1}
%\makeatletter
%\renewcommand{\theequation}{S\arabic{equation}}
%\renewcommand{\thefigure}{S\arabic{figure}}
%\renewcommand{\bibnumfmt}[1]{[S#1]}
%\renewcommand{\citenumfont}[1]{S#1}
%%%%%%%%%% Prefix a "S" to all equations, figures, tables and reset the counter %%%%%%%%%%

%\begin{thebibliography}{99}
%\bibitem{SM-Groot1997} R. Groot and P. Warren, J. Chem. Phys. \textbf{107}, 4423 (1997).

%\end{thebibliography}
\end{document}